\documentclass{article}

\usepackage{amsmath,amssymb}
\usepackage{graphicx}
\usepackage{caption}

\newcommand{\vr}{{\boldsymbol r}}
\newcommand{\F}{\mathcal{F}}
\newcommand{\HH}{\hat{\mathrm{H}}}

\newcommand{\lp}{{\ell_{\textrm p}}}
\newcommand{\lt}{{\ell_{\textrm t}}}
\newcommand{\moyq}[2][]{\left\langle{#2}^2\right\rangle_{#1}}
\newcommand{\kB}{{k_{\textrm B}}}

\newcommand{\un}[1]{\,{\mathrm{#1}}}
\newcommand{\upd}{\,\mathrm{d}}
\newcommand{\Name}[1]{\textrm{#1},}
\newcommand{\Review}[1]{\textit{#1}}
\newcommand{\Vol}[1]{\textbf{#1}}
\newcommand{\Pages}[1]{\textrm{#1}}
\newcommand{\Year}[1]{($\mathrm{#1}$)}
\newcommand{\Page}[1]{\textrm{#1}}
\newcommand{\Book}[1]{\textit{#1}}
\newcommand{\Editor}[1]{\textrm{#1}}
\newcommand{\Publ}[1]{\textrm{#1}}
\newcommand{\Section}[1]{\textrm{#1}}
\newcommand{\SAME}[3]{\textbf{#1} \textrm{#2} ($\mathrm{#3}$)}

\title{DNA loop statistics and torsional modulus}
\author{Vincent Rossetto\\
Laboratoire de physique, \'Ecole normale sup\'erieure de Lyon \\
46 all\'ee d'Italie, 69364 Lyon CEDEX 7, France\footnote{
Present address: Max Planck Institut f\"ur Physik Komplexer Systeme
N\"othnitzer Stra\ss{}e 38, D-01187 Dresden, Germany.}
\footnote{E-mail: \texttt{rossetto@mpipks-dresden.mpg.de}}.}
\begin{document}
\maketitle

\begin{abstract}
The modelling of DNA mechanics under
external constraints is discussed. Two analytical models are
widely known, but disagree for instance on the value
of the torsional modulus. The origin of this embarassing situation
is located in the concept of \emph{writhe}.
This letter presents a unified model for DNA establishing 
a relation between the different approaches.
I show that the writhe created by the loops of DNA is at the origin
of the discrepancy. To take this into account,
I propose a new treatment of loop statistics
based on numerical simulations using the most general
formula for the writhe, and on analytic calculations with
only one fit parameter. 
One can then compute the value of the torsional modulus of DNA
without the need of any cut-off.
\end{abstract}

\section{General motivations}
New experimental techniques in single molecule manipulation of DNA
and supercoiling control have stimulated improvements in the
understanding of DNA mechanics
\cite{brya03,bust94,stri96,stri98}. 
Surprisingly, the measurements have to
be interpreted through a rather sophisticated model in order to extract
physical constants~\cite{mark2,moro98,meza98}. Different approaches
lead to disagreeing values of, in particular, the torsional modulus~$C$
of the molecule along its axis.
The method used by Moroz and Nelson~\cite{moro98}
leads to the value~$C/\kB T=109\un{nm}$,
while using the same experimental data
the model provided by M\'ezard and Bouchiat~\cite{meza98}
gives $84\pm10\un{nm}$.
Recently, another experiment performed two
direct measurements~\cite{brya03} with  a weighted average of
$102\pm6\un{nm}$.
In this letter, I will establish the domains of 
validity of the two mentioned 
theoretical approaches, clarify 
the origin of their disagreement and compute a
value of the torsional constant~$C$ with a new model.

  This letter is organized as follows:
It starts with a short introduction to two widely
used models for elasticity of a polymer.
In order to take into account DNA resistance to torsion, 
one of them, \emph{worm-like chain} model,
is improved with a new elastic energy term.
As these models only contain a \emph{local} description of the molecule,
the following paragraph introduces the global geometrical 
description of a single DNA molecule.
Its three dimensional coiling 
is described by a quantity called \emph{writhe},
whose fluctuations reduce the effective torsional modulus.
I relate the models of Ref.~\cite{moro98}
and \cite{meza98} and study their domains of 
validity. 
The writhe fluctuations
are interpreted in the following paragraph by considering 
the loops formed by the molecule.
The last paragraph describes both a full numerical computation
of the writhe fluctuations and a one parameter fit for the
interpretation in terms of loops. 
The different theoretical models are also compared
to this new treatment of the writhe,
and a value of the torsional modulus is computed.

I will use the notations: $\beta=1/\kB T$,
$F=f\kB T$ the force exerted on the ends of the molecule
and $\theta$ the angle of a vector with the direction of the force,
$L$ the length of the molecule, $A=\lp\kB T$ the bending modulus, 
$C=\lt\kB T$ the torsional modulus,
$\Gamma=\gamma\kB T$ the torque exerted on the free end of the molecule,
$z$ its vertical extension and
$\chi$ the rotation angle of the free end. 
The torsion $\omega(s)$ of the DNA molecule is the difference per
unit length of the
rotation angle of one strand around the other with respect
to the unconstrained state in the same conformation.
It is a function of the arc length~$s$ and can be seen as
``twist angle density'': The torsion integrated along the molecule
is called \emph{twist angle} and is noted $\Omega$.

\section{DNA elasticity models}
Thanks to its double helix structure, DNA is a very stable polymer.
This stability is needed to conserve genetic materials.
It gives opportunity to submit a single molecule to
mechanical constraints (forces above $0.04\un{pN}$)
without destroying it and to measure its response.
The extension $z$ of a single molecule submitted to a force~$F=f\kB T$
is the first quantity that has been studied. It has to be interpreted
within a polymer elasticity model.

The \emph{freely jointed chain} (FJC) describes a polymer as a succession
of independent sticks of length~$b$. For one stick
the elementary partition function is 
\begin{equation}
{\mathfrak z}=\int_{-1}^1 \exp(-fb\cos\theta)\;\upd(\cos\theta)
=\frac{\sinh fb}{fb}.
\label{e.zk}
\end{equation}
For a whole chain, the free energy 
is thus~${\cal F}_{\textrm{FJC}}=\frac{\kB T L}b \ln{\mathfrak z}$,
leading to a relative extension $z/L=fb/3$ when $f$ tends to zero,
and $z/L=1-(fb)^{-1}$ for large forces.

The \emph{worm-like chain} is on contrary a continuous model
based on the bending energy of the axis. 
The state variable is the tangent unit vector,
described by its angle $\theta$ with the force,
thus the phase space is made of functions $\theta(s)$.
The resistance of the molecule to bending creates a
correlation between tangent vectors along a curvilinear
distance~$\lp\simeq50\un{nm}$ called the \emph{persistence length}.
The Hamiltonian for this model is obtained by analogy
with a quantum system (see Ref.~\cite{meza98}):
\begin{equation}
\beta\HH_{\textrm{worm}}(f\lp)=-\frac1{2\sin\theta}\frac\partial{\partial\theta}
\sin\theta\frac\partial{\partial\theta}-f\lp\cos\theta.
\label{e.hw}
\end{equation}
When $f$ tends to zero the
worm-like chain extension is $z/L =\frac23 f\lp$.
In the large force approximation, the molecule is almost
aligned with the force ($\theta\ll1$) and the Hamiltonian~(\ref{e.hw})
reduces to an harmonic oscillator whose ground state gives the free energy 
\cite{mark1}
\begin{equation}
\beta\F_{\textrm{worm}}=\left(-f\lp+\sqrt{f\lp}\right) \frac L\lp
\qquad\qquad\qquad(f\lp\gg1).
\label{e.fver}
\end{equation}
From
$\beta\partial\F_{\textrm{worm}}/\partial(f\lp)=-z/\lp$,
when $f\lp\gg1$, one deduces
\begin{equation}
1-\frac{z}{L}\simeq\frac{1}{2\sqrt{f\lp}}
\qquad\qquad\qquad(f\lp\gg1).
\label{e.zw}
\end{equation}

I now compare the two models.
At low force, entropy dominates and the models both describe an 
object following Hooke's law. 
Requiring the relation $b=2\lp$ between their parameters
makes them equivalent.
This is known to be a good approximation when $f\lp=fb/2\leq1$
or in other terms $F\leq0.08\un{pN}$.
On the contrary at large force, the two models do not
have the same asymptotic behaviour. Comparison between
experimental data and theoretical models 
indicate that the worm-like chain provides a more appropriate
description of DNA elasticity~\cite{bust94}.

The double helix structure of DNA has also for consequence
a resistance to torsion along its axis. 
M\'ezard and Bouchiat~\cite{meza98} showed that taking into account
the local torsion of DNA
introduces new elastic terms in the Hamiltonian~(\ref{e.hw}), 
proportional to
$\gamma^2$: A term due to local torsion energy,
$\beta\HH_{\textrm{torsion}}=-\frac12\gamma^2\, L/\lt$,
and a term related to the geometry of the molecule
\begin{equation}
\beta\HH_{\textrm{rod}}(f\lp,\gamma)=\beta\HH_{\textrm{worm}}(f\lp)+\beta\HH_{\textrm{torsion}}
     -\frac{\gamma^2}2\frac{1-\cos\theta}{1+\cos\theta}.
\label{e.hrod}
\end{equation}
This improved model is called the \emph{rod-like chain}.
In the large force approximation, the denominator~$1+\cos\theta\simeq2$,
thus to second order in~$\theta$ the Hamiltonian rewrites
$\beta\HH_{\textrm{rod}}(f\lp,g)\simeq\beta\HH_{\textrm{worm}}(f\lp-\gamma^2/4)
+\beta\HH_{\textrm{torsion}}-\gamma^2/4$. This approximation is valid
when~$\theta$ is small, thus in a domain where the relation~(\ref{e.fver})
is correct.
One deduces the expression of the free energy of the rod-like chain
in the large force regime:
\begin{equation}
\beta\F_{\textrm{rod}}(f\lp,\gamma)=\beta\F_{\textrm{worm}}(f\lp-\gamma^2/4)-
\frac{1}{2} \gamma^2 \left(\frac{L}{\lt}+\frac{L}{2\lp}\right).
\label{e.relrw}
\end{equation}

\section{The writhe and its fluctuations}
Experimental devices also allow to change the
angle~$\chi$, and to measure how the molecule responds to
this constraint \cite{stri96}. 
One says that the molecule is \emph{supercoiled}.
Supercoiling is quantified by the
writhing angle of the DNA axis~$\mathcal{C}$ \cite{whit89,fain97}:
\begin{equation}
\Phi=\frac{1}{2}\oint_{\mathcal{C}} \upd\vr\cdot
                \oint_{\mathcal{C}} \upd\vr'\times\frac{\vr-\vr'}
{\left\|\vr-\vr'\right\|^3}.
\label{e.calu}
\end{equation}
This expression is \textit{a priori} valid only for {\em closed} chains.
However, in experimental devices the molecule can not go 
around its ends~\cite{stri98},
which is equivalent to having a very long molecule only
manipulated in a small region. The molecule axis can therefore be
imaginarily closed and the 
formula~(\ref{e.calu}) extends to open chains in experimental
conditions~\cite{ross03}. 
Therefore, the fundamental geometrical relation between~$\chi$,
$\Omega$ and $\Phi$ \cite{calu59},
\begin{equation}
\chi=\Omega+\Phi,
\label{th.calu}
\end{equation}
is exact in experimental conditions. 
Formula~(\ref{th.calu}) is widely
known by DNA specialists, in biology and physics~\cite{whit89}.
It explains that a DNA molecule has two ways to deal with
an applied torque : to modify its \emph{local torsion} 
(modify $\Omega$) or to change its shape (modify $\Phi$).
%folding

The twist angle~$\Omega$ is only related to the
torsion. The writhe angle~$\Phi$, as it is computed from
the shape of the axis, is only related to the curvature.
A general study of DNA elasticity performed by Marko and
Siggia~\cite{mark1} asserts that a coupling between
bending and torsion is at least of the third order in strains,
and the estimated coupling constant is small~\cite{moro98}.
As one wants to focus on regimes close to relaxed state,
one shall neglect this coupling and assume that there are
no correlations between~$\Omega$ and~$\Phi$.
(The same assumption has been made in Ref.~\cite{moro98,meza98}.)
Eq.~(\ref{th.calu}) then gives
the fluctuations of $\chi$ in this approximation:
\begin{equation}
\moyq\chi=\moyq\Omega+\moyq\Phi.
\label{e.fcalu}
\end{equation}
The fluctuations of~$\Omega$, thanks to the locality of torsional energy
and independance of $\HH_{\textrm{torsion}}$ with $\theta$,
are proportional to the length~$L$,
$\moyq\Omega=L /\lt$. 
If one measures the torsional modulus of a molecule
only taking the angle~$\chi$ into account, one
one obtains an \emph{effective} value, denoted $\kB T\lt^{\textrm{eff}}$,
related to the real torsional constant, $\kB T \lt$,
through the fluctuations of the writhing angle:
\begin{equation}
\frac{L}{\lt^{\textrm{eff}}}=\frac{L}{\lt}+\moyq\Phi.
\label{e.re}
\end{equation}
The effect of writhing of the molecule is then to reduce the 
measured vale of the torsional modulus between its ends. 
To deduce the value of~$C$ \emph{along the molecule axis}
from measurements, one needs to 
know the writhe fluctuations.

For the rod-like chain model developed in the preceding section,
one remarks that $\moyq\chi$ is deduced from~$\F_{\textrm{rod}}$, 
Eq.~(\ref{e.relrw}), by differentiation
$-2\beta\partial\F_{\textrm{rod}}/\partial(\gamma^2)|_{\gamma=0}\,=\moyq\chi$, 
so
\begin{equation}
\moyq\Phi=\dfrac{1}{2}\left(1-\frac{z}{L}\right)\,\frac{L}{\lp}
\qquad\qquad\qquad(\theta\ll1).
\label{e.mn}
\end{equation} 
This result, combined with relation~(\ref{e.zw}),
was first given by Moroz and Nelson~\cite{moro98} as a correction
like in Eq.~(\ref{e.re}), when the force is large enough: 
\begin{equation}
  \moyq{\Phi}=\frac1{4\sqrt{f\lp}}\,\frac L\lp.
\end{equation}
Their model is therefore an approximation of the more general rod-like chain
model described by M\'ezard and Bouchiat.
Let us now evaluate the
minimum force on their common validity range. Expression~(\ref{e.mn}) is
obtained when the molecule is almost straight, namely when
$f\lp\gg1$.  That might be translated into physical units
to~$F\gg0.1\un{pN}$, thus for forces of the order of~$1\un{pN}$.

While M\'ezard and Bouchiat's approach is more general, it has been 
pointed out that it suffers from a pathology related to the writhe
formulation~\cite{ross02}. More precisely, the expression 
for the writhe used in Ref.~\cite{meza98},
\begin{equation}
\int(1-\cos\theta(s))\upd\phi(s),
\label{e.ful}
\end{equation} 
is equal to formula~(\ref{e.calu}) modulo~$4\pi$ 
($\phi$ is the azimuthal angle of the tangent vector).
They are equal only when the molecule can be straightened out without
cutting it, nor having a point passing through $\theta=\pi$ at any time. 
Otherwise the formul\ae{} differ by a multiple of~$4\pi$ \cite{note1,full78}.
When the molecule is deformed into a blob, such configurations 
where the formul\ae{} disagree are numerous.
Typically, in this case the molecule has \emph{loops}. 
The following section is dedicated to an estimation of the number
of loops in order to extend the validity range of formula~(\ref{e.mn}).

\section{Estimation of the loops contribution}
Each loop contributes to the writhing angle~$\Phi$ by an amount
of the order of one turn. I will note $\Delta^2$ the
mean square value of a loop contribution.
This quantity will be numerically estimated in the next section.
Under the assumption that loops are uncorrelated,
the central limit theorem asserts that the
contribution of loops to~$\moyq{\Phi}$ is $n(f\lp)\Delta^2$, where
$n(f\lp)$ is the average number of loops for~$f\lp$ fixed. 
In Ref.~\cite{moro98}, a treatment of loops was proposed, but
only equilibrium loops were considered. This approach 
does not explicitly provide any loop size.
Since the length of one stick being $2\lp$, the 
model avoids the smaller loops, that require much bending
energy.

The average number of loops is estimated here by defining a loop
as a region where $\cos\theta<0$, which gives, using Eq.~(\ref{e.zk})
and $b=2\lp$,
\begin{equation}
n(f\lp)=\frac Lb\,\frac1{\mathfrak z}\int_{-1}^0 
\exp(-fb\cos\theta)\;\upd(\cos\theta)
\underset{f\to0}{\simeq}\frac L{4\lp}\exp(-f\lp)
. \label{eq.nf}
\end{equation}

Thus one obtains an estimate for~$\moyq\Phi$ in an extended force range
by adding the contribution of the loops (regions where
$\cos\theta<0$), given by the preceding formula,
to the one of the other regions (where it is supposed
that $\theta\ll1$), given by the expression~(\ref{e.mn}):
\begin{equation}
\moyq\Phi\simeq \dfrac12\left(1-\frac zL\right)\frac
L\lp+\frac14\Delta^2\exp(-f\lp)\frac L\lp
\qquad\qquad\qquad(f\lp\leq1).
 \label{eq.T}
\end{equation}
Adding these contributions is allowed here thanks to Fuller's
formula for writhe change during a deformation with fixed boundary conditions
(see Ref.\cite{full78}).
When $f\lp$ is small, the last term of Eq.~(\ref{eq.T}) becomes
dominant: The writhe is dominated by the loops contribution.
$\Delta^2$ is computed in the next section.

\section{Numerical results}
\begin{figure}[tb]
\includegraphics[width=0.9\linewidth]{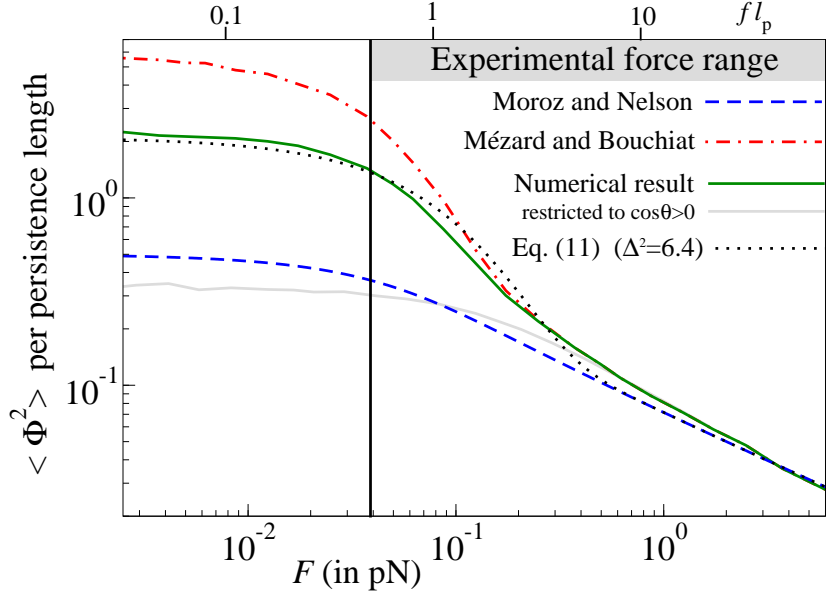}
\caption{Writhe angle fluctuations as a function of the
tension. The experimental force range is indicated by the grey bar.
The plain line presents the results of the present letter model,
compared to modified Moroz and Nelson's result
$\moyq\Phi\lp/L=(1-z/L)/2$ (dashed line) 
and M\'ezard and Bouchiat's (dashed-dotted line). 
Moroz and Nelson's result is valid only for
large forces~($F>1\un{pN}$), validity range of M\'ezard and Bouchiat's 
is wider but fails for the lowest experimental force values,
around~$F=0.2\un{pN}$. The estimated contribution of loops is displayed 
as a dotted line (see text). It corresponds to the best fit for~$\Delta^2$
in Eq.~(\ref{eq.T})~\cite{note2}. 
The grey line is obtained from the same data set, but
after removal of all configurations that have backward facing
regions ; therefore there are no loops in this restricted ensemble.
The results of M\'ezard and Bouchiat's model and the
one presented in this letter have been obtained numerically, 
from equilibrated sets of
50.000 chains of 32~$\lp$ for each point, with discretisation $\lp/30$,
error bars are thiner than the line thickness.
The two other curves are analytical expressions.
\label{f.wr}}
\end{figure}
A Monte-Carlo simulation of semiflexible chains was performed
(described in details in Ref.~\cite{ross03})
and the writhe angle of each configuration of the simulated equilibrated
ensemble was computed, with both expressions~(\ref{e.calu}) and~(\ref{e.ful}).
It has been shown numerically that extra closure
terms are negligible~\cite{ross03}. 
Results are displayed in figure~\ref{f.wr}. The estimate of
Eq.~(\ref{eq.T}) is also displayed for the best value of $\Delta^2=6.4$
(one could conjecture that the geometric exact value be $2\pi$).
The validity ranges estimated
above appear reasonable, and the domains where different
model agree are clearly observed. 
The distribution of writhe angle is {\em Gaussian}, of width
proportional to $L/\lp$ and centered around zero. 
The estimated loops contribution formula agrees quite well with the results
and explains the difference to other models: It is 
not counted in one case and overestimated
in the other one \cite{note2}.
If one removes the curves that have at least one backfacing region,
which corresponds to the definition of a loop (see Eq.~(\ref{eq.nf})),
one gets a curve similar to Moroz and Nelson's.
The three models agree
for $F\geq1\un{pN}$, so the origin of model disagreement is located
at low force.

The numerical results for the writhe angle displayed in figure~\ref{f.wr}
combined with the experimental data of references \cite{stri96,stri98} 
obtained at low force, in the elastic regime, give the value
\[\lt=C/\kB T=93\pm10\un{nm}.\]
This value has to be compared with the ones given
in the introduction. 
The theory of Ref.~\cite{meza98} was applied for forces below this
value, in a domain where it overestimates the writhe fluctuations.
The value obtained by Moroz and Nelson is the closest to
the recent value of Ref.~\cite{brya03}. It was obtained
by eliminating data points obtained with a force lower than
$0.3\un{pN}$ therefore in a domain where all models are equivalent.
It is suggested in Ref.~\cite{brya03} that applying a large
force could lead to an overestimated value for~$C$ 
because of structural modifications of the double helix.
This question is still open until now.

\section{Discussion}
Unlike M\'ezard and Bouchiat's theory, 
neither Moroz and Nelson's model nor this work needs to be 
regularized by any cut-off. 
The value of this cut-off is difficult to relate to
an independently measurable quantity.
It is suggested in Ref.~\cite{moro98} that a treatment of
self-avoidance could be necessary.

In the low force regime, where the freely jointed
chain model is valid for DNA, let us consider one 
molecule as a chain of sticks of length~$b=2\lp$
and diameter $d=2\un{nm}$. The molecule is
located in a region of size $R\sim\sqrt{Lb}$.
In the low force regime, the sticks have
almost random orientations, so
excluded volume interactions can be estimated 
with the second virial coefficient. 
It follows that
excluded volume interactions appear when $\rho^2R^3b^2d\gtrsim1$
where $\rho=(L/b)R^{-3}$ is the density of sticks. 
This gives $L/b\gtrsim (b/d)^2=2500$.
The length of the widely used $\lambda$-phage DNA is
$16\un{\mu m}\simeq 300\lp$\cite{stri96}.
For larger forces, the molecule gets an orientation and self-avoidance 
effects are still smaller.
It is established in Ref.~\cite{mark2} that
self-avoidance effects can be neglected
as long as $z/L\gtrsim0.25$.
This argument does not apply as is in the present study
since DNA stores torsion. It is known, for example
in the case of a plectoneme, that self-avoidance stabilizes
some torque constrained configurations. 
I focused on the elastic regime, where
the torque is small and in presence of a small but
non-zero force, then applying a torque diminishes $z/L$.
In this situation, the assumption that has been made
consists in considering self-avoidance negligible when~$\gamma\to0$,
or quantitatively when $z/L>0.25$.
%For larger torque, this question remains in a large extent unanswered. 
As a consequence of self-avoidance the
statistical weight of conformations with large writhe increases,
in other words~$\Delta^2$ is underestimated in our model.
Consequently, taking into account self-avoidance effects
would result in an small increase of the value of~$C$ given above.

In the numerical work, self-avoidance 
was not taken into account, following the considerations of the preceding
paragraph.
As a consequence, the studied ensemble contains
knotted configurations. I showed from the same numerical
simulations that a $8\lp$ long worm-like chain 
has a probability of $(5\pm1)\,10^{-4}$ to knot~\cite{ross03}.
For a force of $0.02\un{pN}$ the polymer statistic can be split
into independent elastic blobs of size $8\lp$. Then the probability to have
a knot in at least one blob is around $2.10^{-2}$ for a $16\un{\mu m}$
DNA molecule. If the force is higher, the knot probability 
decreases, as blobs become smaller, and is negligible in 
the experimental force range. Knotted configurations are then
believed to play no role in the experimental force range.

\medskip

In this work, I have investigated the connections between different
DNA models and showed that
Moroz and Nelson's model and M\'ezard and Bouchiat's 
are related in a simple way, which depends on how DNA loops are taken
into account.
The estimates of the writhe fluctuations of the molecule in those models
have been compared to numerical results.
I have shown that the differences between those models 
are due to the absence (in Moroz and Nelson's model) 
or the overestimate (in M\'ezard and Bouchiat's model)
of the contribution of the molecule loops to the writhe.
This model takes more accurately writhe fluctuations into account
whithout introducing any cut-off.
It also provides a reasonable value for the torsional
modulus of DNA under small constraints.

\medskip 

\centerline{---}

\medskip

The author would like to thank A. C. Maggs for introduction to this field
and discussions, and ENS-Lyon for financial help.


\begin{thebibliography}{0}
\bibitem{note1}The singularity involved is due to the inexistence
of smooth map from a sphere to a plane, and appears in Euler
angles representation and in Eq.~(\ref{e.hrod}) as well.
\bibitem{note2}The fit for $\Delta^2$
with M\'ezard and Bouchiat's curve is much worse  (not shown).
\end{thebibliography}
\end{document}